\documentclass[aps,prb,twocolumn,superscriptaddress,floatfix]{revtex4}
\usepackage{graphicx}
\usepackage{epsfig}
\usepackage{amsmath}
\usepackage{amssymb}

\begin{document}

\title{Reverse Monte Carlo study of apical Cu--O bond distortions in YBa$_2$Cu$_3$O$_{6.93}$}

\author{Callum A. Young}
\affiliation{Department of Chemistry, University of Oxford, Inorganic Chemistry Laboratory, South Parks Road, Oxford OX1 3QR, U.K.}

\author{Edward Dixon}
\affiliation{Department of Chemistry, University of Oxford, Inorganic Chemistry Laboratory, South Parks Road, Oxford OX1 3QR, U.K.}

\author{Matthew G. Tucker}
\affiliation{ISIS Facility, Rutherford Appleton Laboratory, Harwell Science and Innovation Campus, Didcot OX11 0QX, U.K.}

\author{David A. Keen}
\affiliation{ISIS Facility, Rutherford Appleton Laboratory, Harwell Science and Innovation Campus, Didcot OX11 0QX, U.K.}

\author{Michael A. Hayward}
\affiliation{Department of Chemistry, University of Oxford, Inorganic Chemistry Laboratory, South Parks Road, Oxford OX1 3QR, U.K.}

\author{Andrew L. Goodwin}
\email[]{andrew.goodwin@chem.ox.ac.uk}
\affiliation{Department of Chemistry, University of Oxford, Inorganic Chemistry Laboratory, South Parks Road, Oxford OX1 3QR, U.K.}

\date{\today}
\begin{abstract}
A combination of neutron total scattering measurement and reverse Monte Carlo (RMC) refinement is applied to the study of apical Cu--O bond distortions in the high-$T_{\textrm c}$ superconductor YBa$_2$Cu$_3$O$_{6.93}$. We show that the average structure is not consistent with a split-site model for the corresponding Cu and O positions, but that the local structure nevertheless reveals the existence of two separate apical Cu--O bond lengths. Using $G(r)$ data obtained from a variety of $Q_{\textrm{max}}$ values we show that this result is independent of the data treatment methodology. We also find that the resulting `short' and `long' Cu--O bond lengths agree well with the results of previous EXAFS studies. The existence of bimodal apical Cu--O bond distributions in the context of a single-site average structure model is interpreted in terms of correlated displacements of the Cu and O atoms. We find evidence also for the clustering of short apical Cu--O bonds within our RMC configurations.
\end{abstract}


\maketitle
\section{Introduction}

The mechanistic role of structural inhomogeneities has remained a contentious aspect\cite{Newns_2007,Reznik_2006,Orenstein_2000} of the science of high-temperature superconductors almost since the phenomenon was first discovered---be it in the form of electron-phonon coupling,\cite{Gadermaier_2010} charge localisation (`stripes')\cite{Tranquada_1995} or bipolaron formation and condensation.\cite{Alexandrov_2011} Local-structure probes, such as X-ray absorption spectroscopy and total scattering (or pair distribution function, PDF) measurements, might be expected to be particularly sensitive to the extent and nature of lattice distortions; yet the results of such experiments have often appeared contradictory and remain controversial. As a consequence there exists no generally-accepted microscopic description of local structure distortions in canonical high-$T_{\rm c}$ materials such as the YBa$_2$Cu$_3$O$_{7-\delta}$ (YBCO) family.

\begin{figure}
\begin{center}
\includegraphics{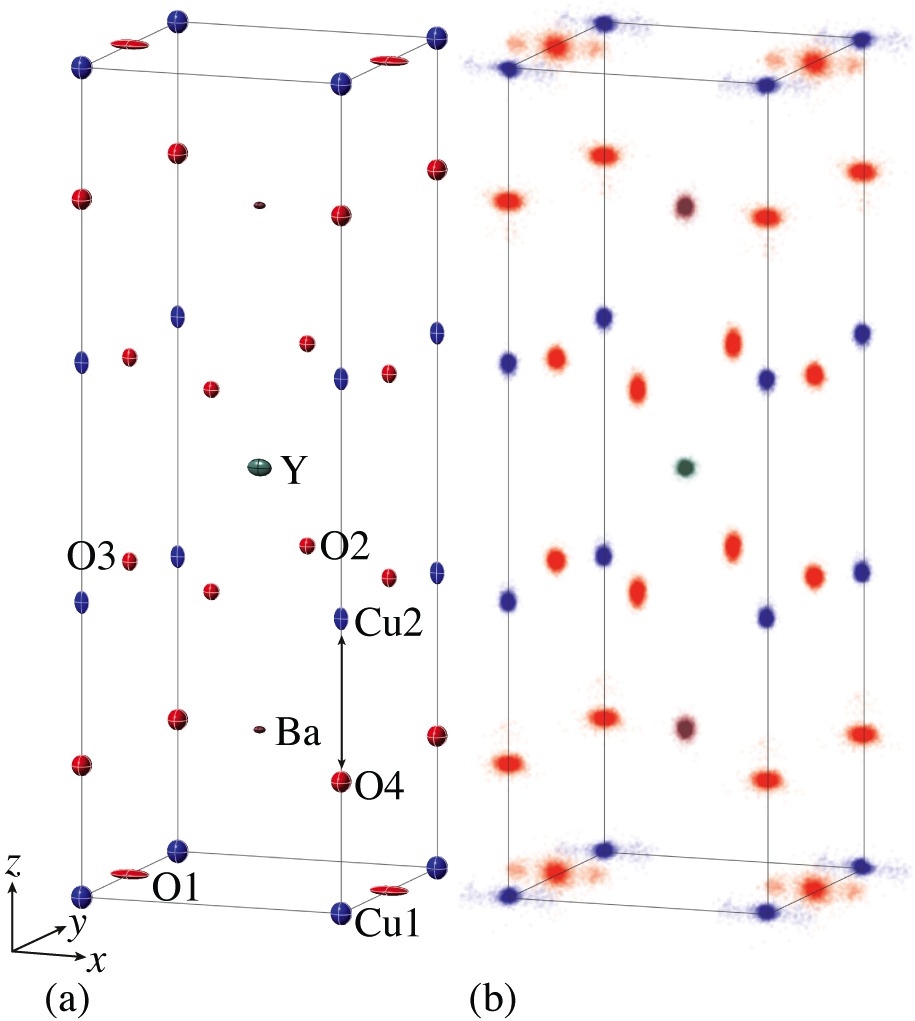}
\caption{Representations of the average structure of YBa$_2$Cu$_3$O$_{6.93}$ determined by (a) Rietveld refinement and (b) RMC refinement of neutron total scattering data. The crucial Cu2--O4 bond is indicated by the arrow. Thermal ellipsoids in (a) are shown at 80\% probability level; the representation in (b) is obtained by projecting the RMC configuration onto a single unit cell.\label{fig1}}
\end{center}
\end{figure}

In the particular case of YBCO, it is the existence of a double-well potential for the apical Cu2--O4 bond that has proven particularly controversial, primarily because of the associated implications for charge localisation and electron-lattice coupling [Fig.~\ref{fig1}(a)].\cite{Jin_2007} An early XANES study was the first to suggest the existence of such a double-well,\cite{Conradson_1989} and a number of EXAFS investigations in the intervening years have essentially supported this same conclusion.\cite{Mustredeleon_1990,Mustredeleon_1992,Stern_1993,Booth_1996,Tyson_1997} In contrast, both single crystal\cite{Sullivan_1993,Schweiss_1994} and powder\cite{Francois_1988,Williams_1988,Kwei_1990,Kwei_1991} diffraction experiments found no evidence for splitting of either the Cu2 or O4 crystallographic sites. Likewise neutron PDF studies ruled out a bimodal distribution of O4 positions but yielded improved fits for models that allowed a variety of different splittings of the Cu2 site.\cite{Louca_1999, Gutmann_2000} And whereas frozen-phonon \emph{ab initio} calculations failed to reproduce a double-well potential for the Cu2--O4 bond,\cite{Liechtenstein_1994}  inelastic neutron scattering measurements linked an increased sensitivity of the dynamic scattering function $S(Q,E)$ near $T_{\textrm c}$ to structural instabilities involving shortening of the same bond.\cite{Arai_1994}

In this paper we attempt to rationalise some of these apparently contradictory results by carrying out a reverse Monte Carlo (RMC) refinement of newly-collected neutron total scattering data for a carefully synthesised YBCO sample of composition YBa$_2$Cu$_3$O$_{6.93}$. This is very much a PDF study, but there are two key differences between our approach and those of Refs.~\onlinecite{Louca_1999} and \onlinecite{Gutmann_2000}. First, where these previous studies have focussed on interpreting structural features of the PDF in terms of increasingly complex single-particle correlations---through symmetry lowering of the unit cell or the introduction of split sites---RMC refinement allows the PDF to be interpreted in terms of \emph{correlated} displacements of atoms. Second, our RMC refinements include simultaneous fits to both PDF and Bragg intensity data, a process that ensures the resultant atomistic configurations are automatically consistent with traditional average-structure refinements. The key result of our study is that the Cu2--O4 bond distribution in our RMC configurations is indeed bimodal, but that correlations between the Cu2 and O4 displacements result in unimodal distributions for both Cu2 and O4 sites. In this way, we are able to rationalise the apparent discrepancy amongst EXAFS, PDF and average structure crystallographic studies.

\section{Materials and Methods}
\subsection{Synthesis and characterisation}

A sample of YBa$_2$Cu$_3$O$_{7-\delta}$ (2.5\,g) was prepared via the citrate gel route described previously.\cite{Blank_1988} A suitable stoichiometric mixture of Y$_2$O$_3$ (99.999\%, dried at 900\,$^{\circ}$C), BaCO$_3$ (99.997\%) and CuO (99.9999\%) was dissolved in a minimum quantity of 6\,M nitric acid. Four mole equivalents of citric acid and 4\,ml of ethylene glycol were then added and the solution heated with constant stirring. The gel formed was subsequently ground into a fine powder, placed in a crucible lined with Ni foil, and heated in air to 600\,$^\circ$C at a rate of 1\,$^{\circ}$C min$^{-1}$. The resulting powder was re-ground and pressed into 13\,mm pellets under a force of 5\,t. All subsequent heating was performed with the sample contained within an alumina boat lined with Ni foil. The sample was fired at 920\,$^{\circ}$C under flowing O$_2$ and then cooled to room temperature at a rate of 1\,$^\circ$C\,min$^{-1}$. This process was iterated three times. After the final cooling cycle, the pellets were re-ground and the resulting powder heated at 600\,$^{\circ}$C for 12\,h under flowing O$_2$ to give a sample with O substoichiometry $\delta=0.07(3)$ as determined by iodometric analysis. Phase purity was confirmed by X-ray powder diffraction; the refined lattice parameters were found to be in good agreement with previously published values.\cite{Sharma_1991} A measurement of the superconducting transition temperature by SQUID magnetometry gave  $T_{\textrm c}=92(1)$\,K.

\subsection{Neutron total scattering}

Neutron total scattering data were collected using the GEM instrument at ISIS.\cite{Williams_1998,Day_2004,Hannon_2005} A sample of polycrystalline YBa$_2$Cu$_3$O$_{6.93}$ (1.8\,g), prepared as described above, was placed within a cylindrical thin-walled vanadium can of 3\,mm diameter and 5.8\,cm height, which was in turn loaded inside a closed cycle helium refrigerator. The sample was cooled to 50\,K and total scattering data then collected over a large range of scattering vectors of magnitudes $0.7\leq Q\leq50$\,\AA$^{-1}$, corresponding to a real-space resolution of order $\Delta r \simeq 3.791/Q_{\textrm{max}} \simeq 0.08$\,\AA.

Following their collection, the total scattering data were corrected using standard methods, taking into account the effects of background scattering, absorption, multiple scattering within the sample, beam intensity variations, and the Placzek inelasticity correction.\cite{Dove_2002} These corrected data were then converted to experimental $F(Q)$ and $G(r)$ functions:\cite{Dove_2002,Keen_2001}
\begin{eqnarray}
F(Q)&=&\rho_0 \int_0^\infty4\pi r^2 G(r)\frac{\sin Qr}{Qr}\mathrm{d}r\\
G(r)&=&\sum_{i,j=1}^nc_ic_j\bar{b}_i\bar{b}_j[g_{ij}(r)-1],
\end{eqnarray}
where
\begin{equation}
g_{ij}(r)=\frac{n_{ij}(r)}{4\pi r^2\mathrm{d}r\rho_j},
\end{equation}
$n_{ij}(r)$ the number of pairs of atoms of type $i$ and $j$ separated by distance $r$, $\rho_0$ is the number density, $c_i$ the concentration of each species $i$ and $b_i$ the corresponding neutron scattering length. As part of the data normalisation process it is usual practice to determine a useable value $Q_{\textrm{max}}$ that represents the best possible compromise between optimising $\Delta r$ for PDF refinement (\emph{i.e.} making $Q_{\textrm{max}}$ as large as possible) and avoiding the unnecessary inclusion of high-frequency noise that can plague high-$Q$ data. In this instance, we found that a value $Q_{\textrm{max}}=40$\,\AA$^{-1}$ produced the most reliable $G(r)$ function.

The Bragg profile functions for each data set were extracted from the scattering data collected by the detector banks centred on scattering angles $2\theta = 54.46^{\circ}$, $63.62^\circ$ and $91.37^{\circ}$.
The experimental Bragg diffraction profiles were fitted with the {\sc gsas} Rietveld refinement program\cite{GSAS} using the published structural model;\cite{Capponi_1987}  the fitting process is discussed in greater detail in the Results section below.

\subsection{Reverse Monte Carlo refinement}

The reverse Monte Carlo refinement method as applied to crystalline materials, together with its implementation in the program {\sc rmcprofile} have been described in detail elsewhere.\cite{Dove_2002,Tucker_2007} The basic refinement objective is to produce large atomistic configurations that can account simultaneously for the experimental $F(Q)$, $G(r)$ and Bragg profile $I(t)$ functions. This is achieved by accepting or rejecting random atomic moves subject to the metropolis Monte Carlo algorithm, where in this case the Monte Carlo acceptance criterion is determined by the quality of the fits to data. The refinement process is continued until no further improvements in the fits to the data are observed. By virtue of the particular importance in the present study of fitting accurately the lowest-$r$ region of the PDF---which includes the all-important distribution of Cu2--O4 separations---we applied a larger weighting to the region $0\leq r\leq7$\,\AA\ than to the remainder of the PDF.

Our starting configurations for the RMC process were based on a $24\times24\times8$ supercell of the crystallographic unit cell shown in Fig.~\ref{fig1}(a). Each configuration contained 59\,904 atoms and extended approximately 90\,\AA\ in each direction. By virtue of the small experimental value of $\delta$ for our sample, and in order to avoid introducing model bias, we did not incorporate any explicit consideration of  O vacancies within our model. In addition to the required fits to data, the only constraints placed on the atomic coordinates were a set of `distance window' constraints which act to maintain an appropriate framework connectivity throughout the refinement process.\cite{Tucker_2007,Goodwin_2005} In each case the values used [see Table~\ref{table1}] were based on the extrema of the corresponding peaks in the experimental PDF.

\begin{table}
\caption{\label{table1} `Distance window' parameters \emph{d}$_{\textrm{min}}$, \emph{d}$_{\textrm{max}}$ used for all RMC refinements in this study.}
\begin{center}
\begin{tabular}{@{\extracolsep{5mm}}lcc}      
\hline\hline Atom pair&$d_{\textrm{min}}$ (\AA)&$d_{\textrm{max}}$ (\AA)\\\hline
Y--O2,3&2.10&2.56\\
Ba--O&2.50&3.50\\
Cu1--O1&1.73&2.10\\
Cu1--O4&1.73&2.50\\
Cu2--O2&1.73&2.10\\
Cu2--O3&1.73&2.10\\
Cu2--O4&1.73&2.50\\\hline\hline
\end{tabular}
\end{center}
\end{table}
 
\section{Results}

\subsection{Average structure determination}

\begin{figure}
\begin{center}
\includegraphics{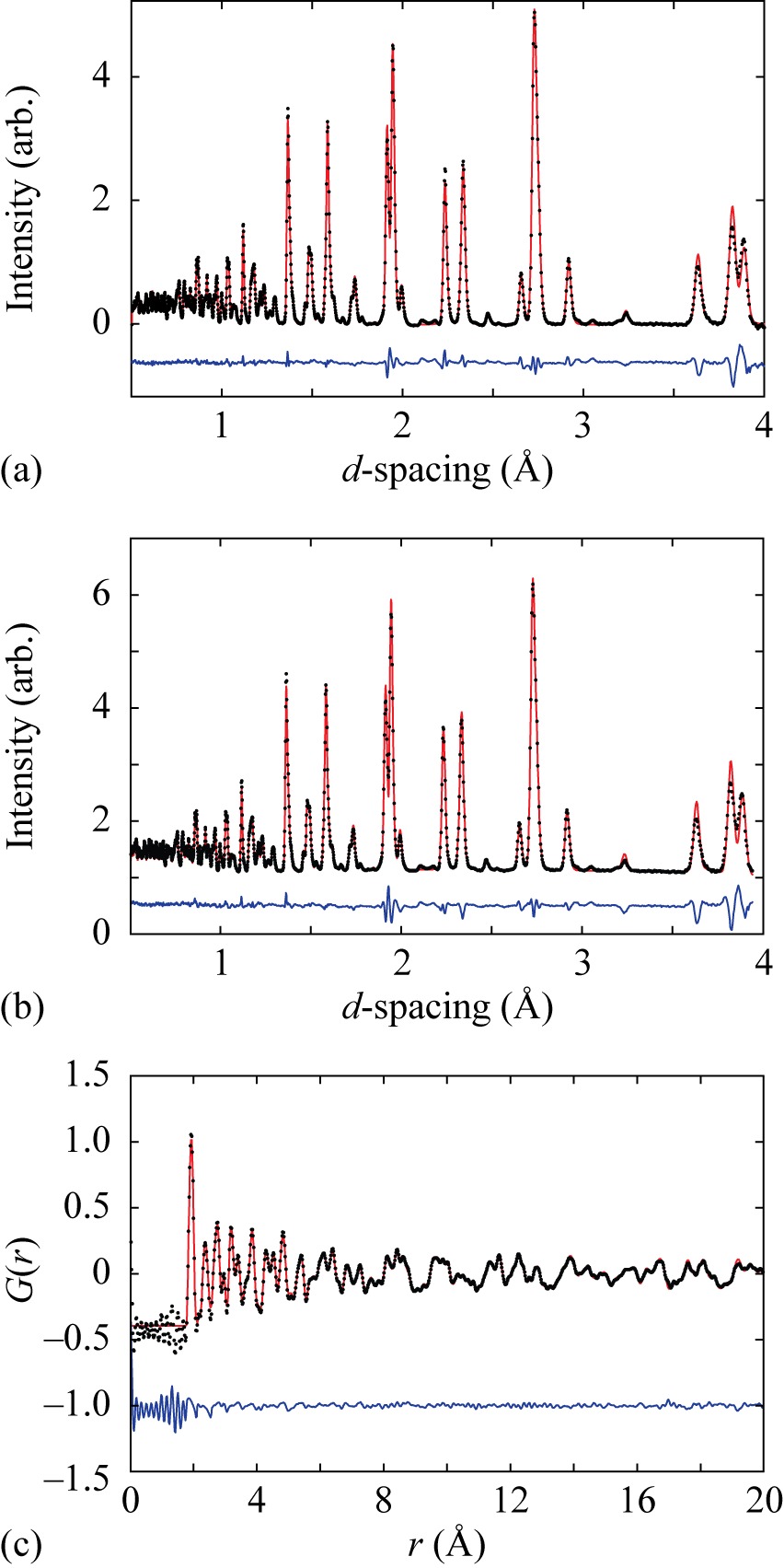}
\caption{The fit to neutron powder diffraction data obtained by (a) Rietveld and (b) RMC refinement. (c) The corresponding $G(r)$ fit obtained using RMC. In all cases, data are shown as filled circles, fits as red lines and the difference (data$-$fit) as blue lines, sometimes shifted vertically for clarity. A refined background function has been subtracted from the data and fit shown in (a).\label{fig2}}
\end{center}
\end{figure}

Our first step was to use the {\sc gsas} refinement package\cite{GSAS} to verify that the Bragg contribution to our total scattering data could be interpreted in terms of sensible lattice parameters, atomic coordinates and anisotropic displacement parameters. The values obtained, which are summarised in Table~\ref{table2}, are in good agreement with previous crystallographic studies of nearly-stoichiometric YBCO samples.\cite{Capponi_1987,Kwei_1990,Schweiss_1994} The corresponding fit to data is shown in Fig.~\ref{fig2}(a) and a representation of the structural model obtained is that shown in Fig.~\ref{fig1}(a). It was possible to achieve stable refinement of anisotropic displacement parameters for all atoms other than Ba, for which unconstrained refinement consistently yielded non-positive definite values for the $U_{22}$ parameter; consequently this value was fixed at 0.001\,\AA$^2$. We remark also that displacements of the so-called `chain' oxygen atoms (O1) are clearly very anisotropic. This behaviour, which has been reported elsewhere,\cite{Gutmann_2000,Sharma_1991,Schweiss_1994} is thought to be related to low-energy correlated displacements of the Cu1-centred [CuO$_4$] square units.

\begin{table}
\begin{center}
\caption{\label{table2}Crystallographic parameters, atomic coordinates and isotropic equivalent displacement parameters determined using Rietveld refinement of neutron scattering data for YBa$_2$Cu$_3$O$_{6.93}$.}
\begin{tabular}{@{\extracolsep{5mm}}lcccc}
\hline\hline
\multicolumn{1}{l}{Crystal system}&\multicolumn{4}{l}{Orthorhombic}\\
\multicolumn{1}{l}{Space group}&\multicolumn{4}{l}{$Pmmm$}\\
\multicolumn{1}{l}{$a$ (\AA)}&\multicolumn{4}{l}{3.81412(5)}\\
\multicolumn{1}{l}{$b$ (\AA)}&\multicolumn{4}{l}{3.87694(6)}\\
\multicolumn{1}{l}{$c$ (\AA)}&\multicolumn{4}{l}{11.63970(22)}\\
\multicolumn{1}{l}{$V$ (\AA$^3$)}&\multicolumn{4}{l}{172.1170(30)}\\
\multicolumn{1}{l}{$Z$}&\multicolumn{4}{l}{1}\\
\multicolumn{1}{l}{$T$ (K)}&\multicolumn{4}{l}{50}\\
\hline\hline
Atom&$x$&$y$&$z$&$U_{\rm iso}$ (\AA$^2$)\\\hline
Y&0.5&0.5&0.5&0.0031(4)\\
Ba&0.5&0.5&0.18377(12)&0.00092(26)\\
Cu1&0&0&0&0.0038(4)\\
Cu2&0&0&0.35535(9)&0.0019(8)\\
O1&0&0.5&0&0.0115(7)\\
O2&0.5&0&0.37837(13)&0.0045(4)\\
O3&0&0.5&0.37727(15)&0.0020(4)\\
O4&0&0&0.15879(11)&0.0047(4)\\\hline\hline
\end{tabular}
\end{center}
\end{table}

Since our study concerns primarily the distribution of Cu2--O4 bonds, we investigated the validity of structural models which incorporated splitting of one or other or both of the two atomic sites. As in previous Rietveld studies,\cite{Francois_1988,Williams_1988,Kwei_1990,Kwei_1991} we found that none of these three cases yielded a stable refinement.

\subsection{Local structure determination}

RMC refinement yielded equally satisfactory fits to the neutron scattering data---both in terms of the reciprocal-space Bragg intensity function $I(t)$ and the real-space $G(r)$ transform. Importantly, the quality of the Bragg profile fit is comparable to that obtained from Rietveld refinement, implying that the RMC configurations are as consistent with the Bragg component to the total scattering function as the average-structure model given in Table~\ref{table2}.

By collapsing the atomic coordinates of an entire RMC configuration onto a single unit cell, it is possible to visualise the RMC-refined average structure [Fig.~\ref{fig1}(b)]. As found in the Rietveld model, the O1 distribution is anisotropic (and indeed there is some sign here of a split site, although we do not pursue this possibility further here). Encouragingly, the Ba site is now represented by a physically sensible distribution function; because RMC configurations represent a physical box of atoms, there is never an issue of obtaining non-positive-definite anisotropic displacement parameters.

\begin{figure}
\includegraphics{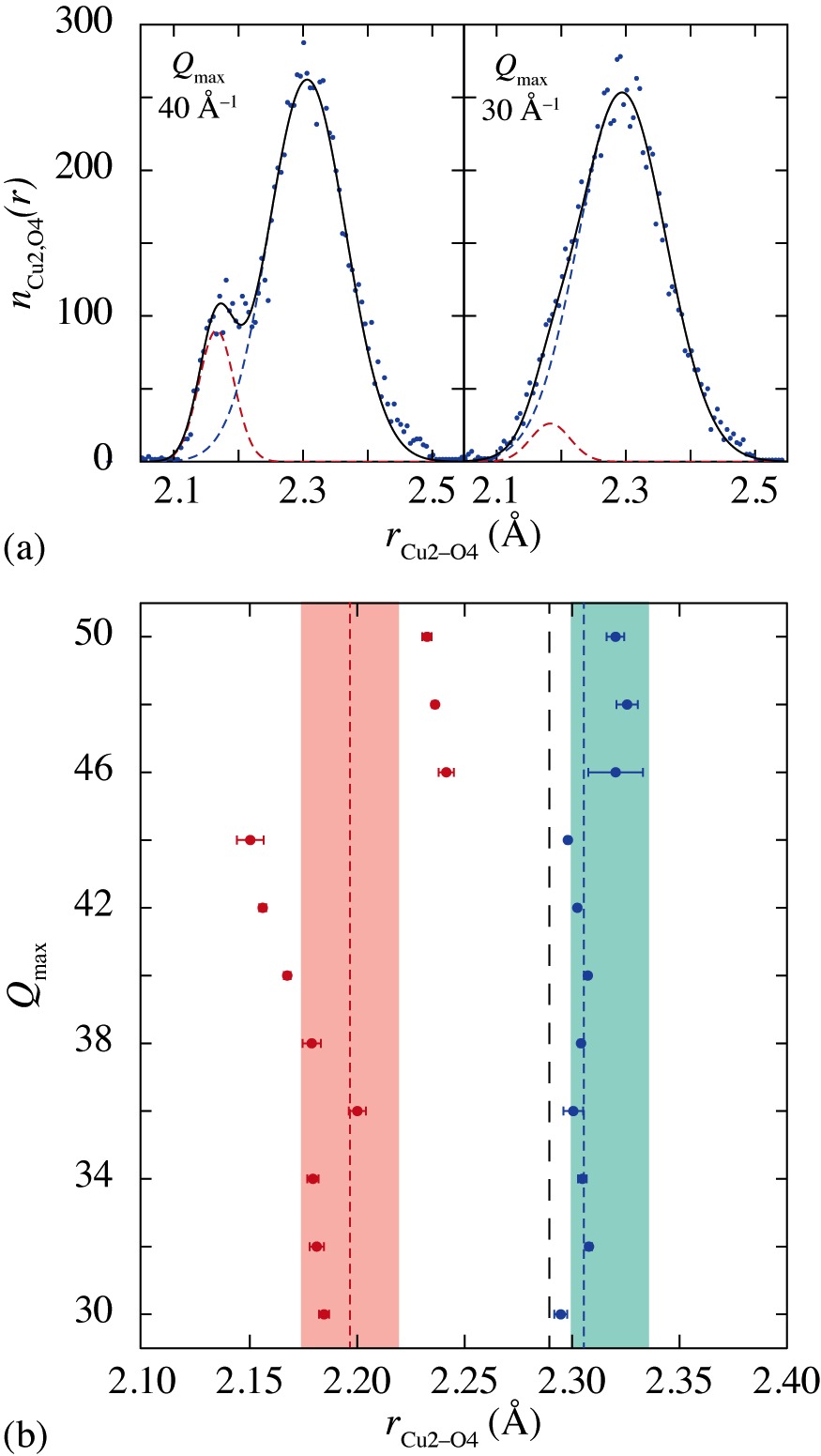}
\caption{\label{fig3}(a) Cu2--O4 partial pair distribution functions (solid points) determined by RMC refinements against $G(r)$ functions generated for maximum scattering vectors of (left) $Q_{\textrm{max}}=40$\,\AA$^{-1}$ and (right) $Q_{\textrm{max}}=30$\,\AA$^{-1}$. These two distributions correspond respectively to the most and the least convincing bimodal distributions obtained for the 11 different $Q_{\textrm{max}}$ values investigated. Calculated fits using a pair of Gaussian curves are shown as a solid line, with the two individual Gaussian contributions shown as red and blue dashed lines. (b) $Q_{\textrm{max}}$-dependence of the `short' and `long' Cu2--O4 bond lengths (red and blue circles, respectively), determined as the midpoints of the corresponding Gaussian fits. Error weighted averages are plotted as vertical red and blue dashed lines. The range of values obtained using EXAFS in Refs.~\onlinecite{Mustredeleon_1992} and \onlinecite{Booth_1996} are shown as shaded regions, with the Rietveld Cu2--O4 bond length value shown as a bold black vertical line. }
\end{figure}

Because the RMC-refined distributions of atomic positions are not constrained to assume ellipsoidal forms, we are able to check in a straightforward and unbiased manner whether or not there is any real evidence for splitting on either the Cu2 or O4 sites. In both instances a projection of the corresponding atomic coordinates onto the $z$ axis gave distributions that could be well fitted using single Gaussian functions. Hence we can say that our RMC refinements are consistent with unimodal single-particle correlation functions for both atom sites. In particular, there can be no need to invoke split Cu2/O4 atom sites in order to fit the neutron scattering data since RMC would have been free to do so should the data have demanded.

With the RMC description of average structure in YBa$_2$Cu$_3$O$_{6.93}$ established, we turned our attention to the pair correlations at the heart of the local structure controversy. It is straightforward to calculate from an RMC configuration the distribution functions for specific bonding interactions. Carrying out such a calculation for the crucial apical Cu2--O4 bonds yields a distribution that is quite obviously bimodal [Fig.~\ref{fig3}(a), left-hand panel]. The distribution is will fitted using a sum of two Gaussian components, ascribable to contributions from `short' and `long' Cu2--O4 bonds. The midpoints of these Gaussian fits suggested a pair of Cu2--O4 bond lengths that were surprisingly similar to the values obtained in the EXAFS study of Ref.~\onlinecite{Booth_1996}: $r_{\textrm{Cu2--O4}}=2.167$ and 2.307\,\AA\ (RMC) \emph{vs} $r_{\textrm{Cu2--O4}}=2.220$ and 2.337\,\AA\ (EXAFS). 

\begin{figure}
\includegraphics{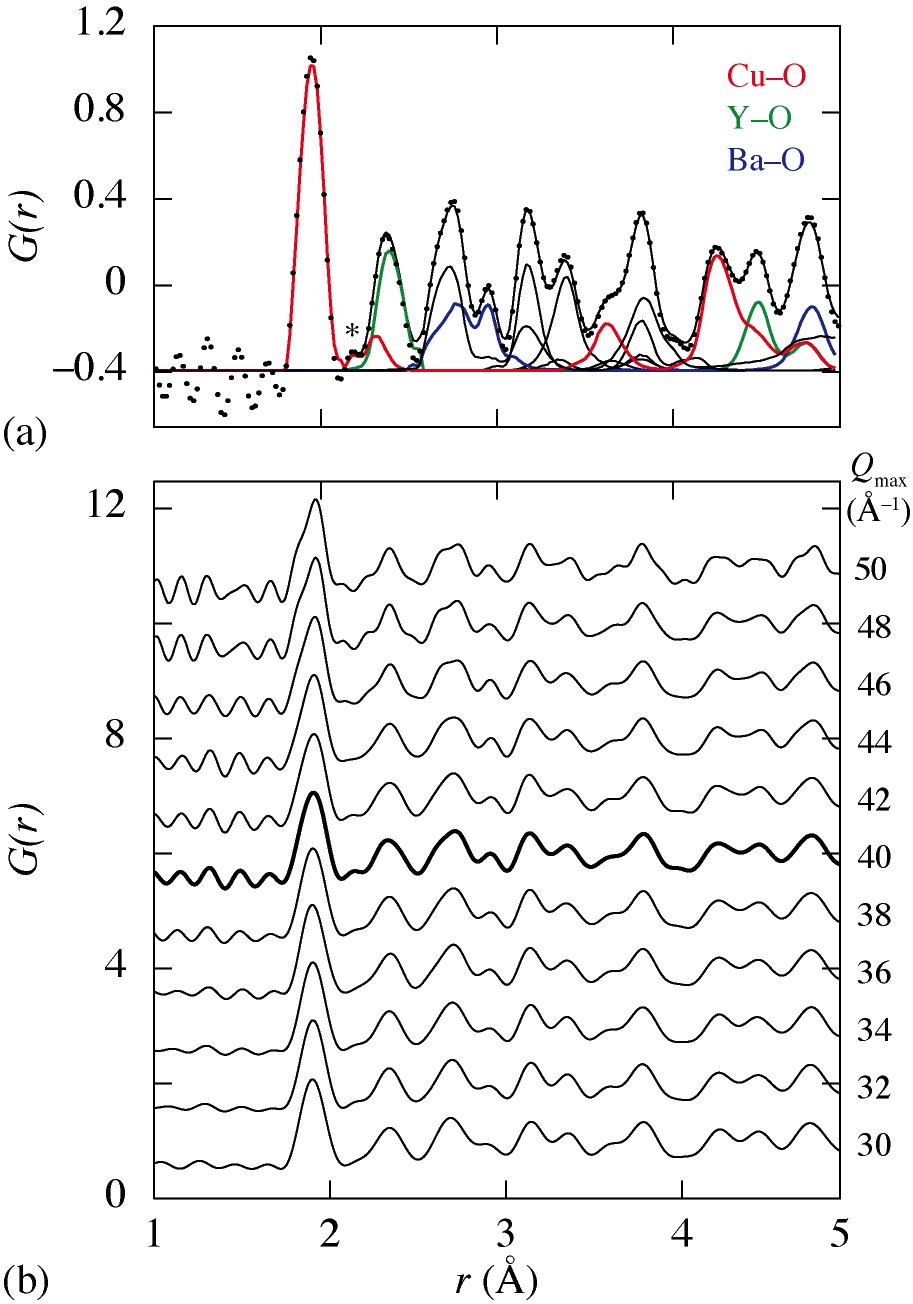}
\caption{\label{fig4} (a) Experimental $G(r)$ function (filled circles) and corresponding RMC fit (solid black line). The relative contributions to the RMC fit arising from the various partial pair distribution functions are included, with those corresponding to the Cu--O, Y--O and Ba--O pairs highlighted in red, green and blue, respectively. The experimental feature responsible for splitting the Cu2--O4 distribution is highlighted with an asterisk. (b) Experimental $G(r)$ functions generated using a range of $Q_{\textrm{max}}$ values $30\leq Q_{\textrm{max}}\leq50$\,\AA$^{-1}$; the bold curve is that shown in (a).}
\end{figure}

We were interested to establish the origin of this feature in the experimental $G(r)$ function in order to determine the extent to which the close agreement with EXAFS results is fortuitous. There is a clear peak in the $G(r)$ function responsible for splitting the Cu2--O4 distribution [Fig.~\ref{fig4}(a)]. The only non-zero partial pair distribution function in this particular region is $g_{\textrm{Cu2,O4}}(r)$. All other Cu--O bonds are much shorter and contribute to the first significant peak seen at $r\simeq1.9$\,\AA. There is some overlap at higher values of $r$ between the Cu2--O4 and Y--O2/Y--O3 partial pair distribution functions, but the values of $g_{\textrm{Y,O2}}(r)$ and $g_{\textrm{Y,O3}}(r)$ are essentially zero for $r<2.20$\,\AA, despite the RMC constraints allowing Y--O distances to be as small as 2.10\,\AA. Consequently the $r\simeq2.15$\,\AA\ feature in the $G(r)$ function---if real---can be ascribed only to a separate distribution of short Cu2--O4 bonds, and hence is an experimental indication of the existence of a bimodal Cu2--O4 bond distribution.

However, the key question here is whether or not the feature is indeed real---our concern being that its magnitude is comparable to the size of the Fourier truncation ripples evident at the lowest values of $r$. Consequently, we proceeded to check whether or not the existence of a bimodal distribution was affected by the value of $Q_{\textrm{max}}$ used to generate the $G(r)$ function. Changing $Q_{\textrm{max}}$ varies strongly the location and periodicity of Fourier ripples but ought not greatly affect the positions and intensities of any real features of the data (though one must keep in mind that reducing $Q_{\textrm{max}}$ also reduces the real-space resolution $\Delta r$). To this end we calculated $G(r)$ data for 11 equally-spaced values $30\leq Q_{\textrm{max}}\leq50$\,\AA$^{-1}$ [Fig.~\ref{fig4}(b)] and used each of these modified data sets to drive an independent RMC refinement. On convergence, refinement was continued for sufficiently many RMC steps to obtain a minimum of seven independent configurations for each run (independent configurations being separated by $140\,000\simeq59\,904\ln10$ accepted moves\cite{Goodwin_2005}).

\begin{table*}
\begin{center}
\caption{Free atom position parameters and Cu2--O4 bond lengths determined using various experimental techniques and refinement strategies.\label{table3}}
\begin{tabular*}{\textwidth}{@{\extracolsep{\fill}}lcccccc}
\hline\hline
&\multicolumn{2}{c}{Rietveld}&\multicolumn{2}{c}{PDFGui}&RMC&EXAFS\\
&this work&Ref.~\onlinecite{Schweiss_1994}&this work&Ref.~\onlinecite{Gutmann_2000}&this work&Ref.~\onlinecite{Booth_1996}\\
\hline
$z_{\textrm{Ba}}$&0.18377(12)&0.18369(12)&0.18357(93)&0.1840(2)&0.18398(9)&--\\
$z_{\textrm{Cu2}}$&0.35535(9)&0.35464(8)&0.3570(49)&0.3548(2)&0.35580(7)&--\\
$z_{\textrm{O2}}$&0.37837(13)&0.37808(7)&0.3778(29)&0.3780(3)&0.37701(11)&--\\
$z_{\textrm{O3}}$&0.37727(15)&0.37808(7)&0.378(2)&0.3782(3)&0.37805(8)&--\\
$z_{\textrm{O4}}$&0.15879(11)&0.15919(10)&0.15998(95)&0.1599(2)&0.15951(9)&--\\\hline
$\langle d($Cu2--O4$)\rangle$&2.2879(17)&2.2860(15)&2.268(14)&2.280(3)&2.2849(13)&--\\
$d($Cu2--O4$)_{\textrm{short}}$&--&--&2.20(10)&2.19(3)&2.197(10)&2.220(5)\\
$d($Cu2--O4$)_{\textrm{long}}$&--&--&2.29(10)&2.37(3)&2.306(3)&2.337(5)\\\hline\hline
\end{tabular*}
\end{center}
\end{table*}

For each of the 11 sets of RMC configurations produced in these refinements, the corresponding Cu2--O4 bond-length distributions were calculated and fitted as described above. It was found that two Gaussian functions were \emph{always} required for these fits and that the corresponding Gaussian midpoints did not vary particularly strongly with $Q_{\textrm{max}}$ [Fig.~\ref{fig3}(c)]. The values of the short and long Cu2--O4 bond lengths we obtain as averages over the various configurations are compared with the EXAFS results of Ref.~\onlinecite{Booth_1996} in Table~\ref{table3}. To give an idea of the level of confidence in these values, we have shown examples of the distributions corresponding to both the most and the least convincing bimodal fits [Fig.~\ref{fig3}(a)]. There is an additional consistency in the Gaussian widths and peak areas: in both cases the values obtained across all 11 data sets vary by less than $10$\%. Taking an error-weighted average across all 11 sets of configurations, we obtain a fraction of short Cu2--O4 bonds of 12.3$\pm$2.8\%. For completeness, we note that it is not at all unexpected that the bond lengths extracted for the lowest and highest $Q_{\textrm{max}}$ values are the least reliable: in the former case, the real-space resolution $\Delta r$ is heavily reduced, and in the latter case the incorporation of high-$Q$ noise begins to affect noticeably the smoothness of the $G(r)$ function.

\subsection{PDFgui refinements}

In order to ensure consistency with previous PDF studies, we performed a final set of PDF refinements using the {\sc pdfgui} software package.\cite{Farrow_2007} Sometimes termed a `real-space Rietveld' approach, {\sc pdfgui} uses the experimental $G(r)$ function to refine atomic coordinates and displacement parameters within a single unit cell. Our aim here was to reproduce the results of Ref.~\onlinecite{Gutmann_2000}: namely, that a split site could be refined for the Cu2 atom but not for the apical O4 atom, and that the split Cu2 atom sites corresponded to a difference $\Delta d$(Cu2--O4) between `short' and `long' Cu2--O4 bond lengths of 0.18(6)\,\AA.

\begin{figure}
\begin{center}
\includegraphics{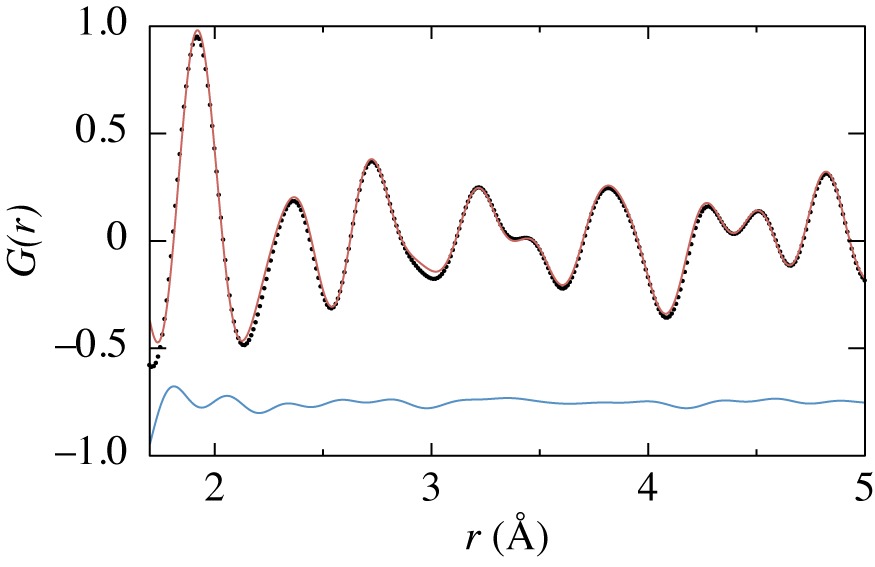}
\caption{A typical fit to $G(r)$ data obtained using {\sc pdfgui}, in this instance for the experimental $G(r)$ function obtained using $Q_{\textrm{max}}=25$\,\AA$^{-1}$. \label{fig5}}
\end{center}
\end{figure}

Our {\sc pdfgui} refinements yielded the fit shown in Fig.~\ref{fig5}, where we have used a $Q_{\textrm{max}}$ value of 25\,\AA$^{-1}$ in order to replicate the conditions of Ref.~\onlinecite{Gutmann_2000}. It was noted in this earlier study that the range of $r$ values included in the fitting process had a significant effect on the value of $\Delta d$(Cu2--O4) obtained; this $r$-dependency in {\sc pdfgui} refinements is symptomatic of the existence of short-range correlations that are not well described by the long-range structural periodicity (see \emph{e.g.}\ Ref.~\onlinecite{Bozin_2007}). Consequently, we determined values of $\Delta d$(Cu2--O4) for the two values employed in Ref.~\onlinecite{Gutmann_2000} ($r_{\textrm{max}}=5$ and 15\,\AA) and also for a range of $G(r)$ functions generated using increasingly large $Q_{\textrm{max}}$ values; our results are given in Tables~\ref{table3} and \ref{table4}.

While we found that it was possible to refine split sites for the Cu2 atom, the difference between `long' and `short' Cu2--O4 bonds was slightly smaller than that reported in Ref.~\onlinecite{Gutmann_2000}. Like the earlier study, we also found that split-site refinements gave improved fits when only the lowest-$r$ region of the PDF was used, suggesting that the effects of site splitting only impact noticeably on the PDF only over short distances. We note that, for some of the various combinations of $Q_{\textrm{max}}$ and $r_{\textrm{max}}$, it was actually possible to refine a split O4 site; however the corresponding distance between sites was never larger than 0.05\,\AA.

\begin{table}
\begin{center}
\caption{Differences $\Delta d$(Cu2--O4) between short and long Cu2--O4 bond lengths determined using $r$-dependent {\sc pdfgui} refinement of the experimental $G(r)$ functions generated for various $Q_{\textrm{max}}$ values.\label{table4}}
\begin{tabular*}{\columnwidth}{@{\extracolsep{\fill}}lcccc}
\hline\hline
$Q_{\textrm{max}}$&$r_{\textrm{max}}$&$\Delta d$(Cu2--O4)&$R_{ws}$&$R_{wu}$\\
 (\AA$^{-1}$)&(\AA)&(\AA)&&\\
\hline
25&5&0.08(7)&0.0908&0.0935\\
&15&0.03(5)&0.0883&0.0901\\
30&5&0.089(26)&0.1224&0.1406\\
&15&0.03(6)&0.0941&0.0951\\
40&5&0.099(28)&0.1207&0.1267\\
&15&0.065(26)&0.1261&0.1276\\
50&5&0.093(22)&0.1863&0.1888\\
&15&0.008(4)&0.2146&0.2157\\
\hline\hline
\end{tabular*}
\end{center}
\end{table}

\section{Discussion}

Perhaps the key result of our RMC study is to demonstrate that a single atomistic configuration can account at once for a number of the seemingly-disparate experimental results reported previously forYBa$_2$Cu$_3$O$_{7-\delta}$:
\begin{enumerate}
\item{The finding from both single crystal and powder diffraction studies that the scattering distributions at the Cu2 and O4 sites of the average structure are unimodal.}
\item{That two Cu2--O4 bond lengths are required to obtain satisfactory fits to Cu $K$-edge EXAFS spectra.}
\item{For {\sc pdfgui} refinements over small regions in real space ($r_{\textrm{max}}\simeq5$\,\AA$^{-1}$), neutron PDF data are best fitted in terms of a structural model with two Cu2 sites (and hence two Cu2--O4 distances); however as $r_{\textrm{max}}$ is increased the refined splitting between Cu2 sites vanishes.} 
\end{enumerate}
The clear implication is that these results are not contradictory, but merely reflect the importance of taking into account length scale when interpreting structural features of YBa$_2$Cu$_3$O$_{7-\delta}$.

Consistency between a bimodal Cu2--O4 bond length distribution on the one hand, and single Cu2/O4 sites in the average structure on the other hand, demands that the Cu2 and O4 displacements are correlated during short- and long-bond formation. Specifically, shortening of the Cu2--O4 bond must involve cooperative displacement of the Cu2 atom in the $-z$ direction [with reference to the coordinates and axes of Fig.~\ref{fig1}(a)] and of the O4 atom in the $+z$ direction; lengthening of the same bond must involve correlated displacement in the opposite direction. A split-site model for Cu2 and O4 positions does not capture this situation properly in either average-structure or PDF refinements. In the former case the model fails because displacements away from the average Cu2/O4 sites will be indistinguishable from thermal motion if correlation is not taken into account;\cite{uijnote} in the latter case---and again because correlation is ignored---the model actually predicts a trimodal Cu2--O4 distribution, which is clearly inconsistent with the bimodal distribution observed experimentally.

For the low-$r_{\textrm{max}}$ {\sc pdfgui} refinements, splitting of the Cu2 site allows the bimodal Cu2--O4 bond distribution to be fitted, but does so with a model that is incompatible with the average structure; hence if the same model is fitted over larger regions in real-space then either the fit is degraded or, if the magnitude of the Cu2 splitting is allowed to refine, a single-site model is recovered. The key advantage of supercell refinement techniques such as RMC is that the large box size allows sufficient sampling to capture both local and periodic structural features within a single atomic-scale model.

Certainly our RMC refinements suggest that in YBa$_2$Cu$_3$O$_{6.93}$ there are two types of apical Cu2--O4 bond and, as already discussed, the difference in bond length that we measure (0.11\,\AA) is remarkably consistent with the results of previous EXAFS studies. There will of course be a temptation to associate with the two Cu2--O4 bond types a different valence state for the Cu2 atom; however we note only that the neutron PDF measurements we have performed here involve no scattering contribution from the electronic structure of the material and hence we do not comment further on this aspect.

\begin{figure}
\begin{center}
\includegraphics{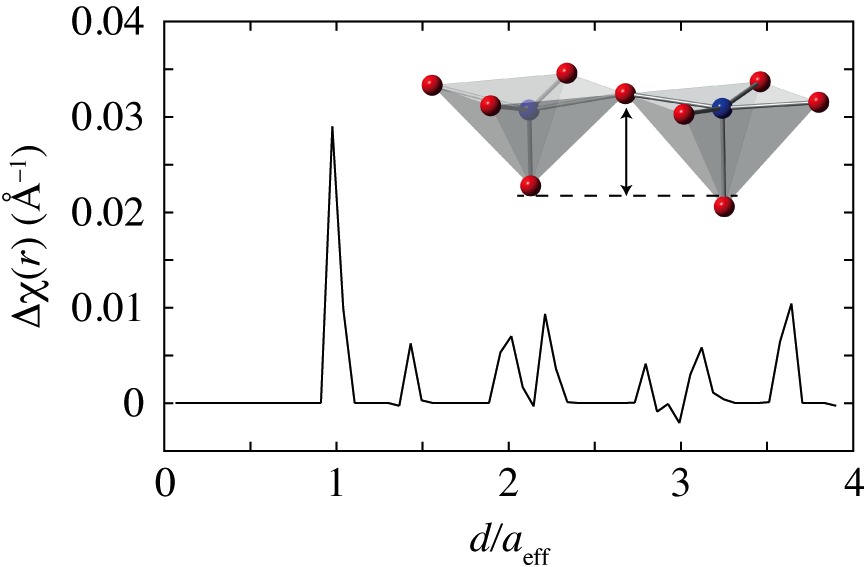}
\caption{\label{fig6}The short-bond/short-bond correlation function calculated for one of the RMC configurations described in the text. For ease of interpretation, the horizontal axis corresponds to interatomic distances scaled relative to the average in-plane lattice parameter. The large positive peak at $r/a_{\textrm{eff}}$ indicates that short apical Cu2--O4 bonds are likely to form neighbouring pairs.}
\end{center}
\end{figure}

We might, however, expect our RMC configurations to be sensitive to correlations in the spatial arrangement of the different types of Cu2--O4 bonds. We have been made aware of an earlier, unpublished, RMC study in which short-bond/short-bond correlations are discussed,\cite{Alexandrov_1998} and we conclude here with a similar calculation for our own RMC configurations. In order to quantify the extent to which short Cu2--O4 bonds are found to cluster, we define a correlation function
\begin{equation}
\chi(r)=\frac{1}{Nr}\sum_i\sum_ju_iu_j(r),\label{eqn}
\end{equation}
where
\begin{equation}
u_i=\left\{\begin{array}{ll}+1\quad&d({\textrm{Cu2--O4}})_i>d_{\textrm c}\\[5pt]-1\quad&d({\textrm{Cu2--O4}})_i<d_{\textrm c}\end{array}\right.,
\end{equation}
the distance $d_{\textrm c}$ is the critical Cu2--O4 bond length below which a bond is considered to be `short', $N$ is the number of Cu2 centres in a configuration, and the sums in Eq.~\ref{eqn} are taken over all Cu2 centres $i,j$ within the same [CuO$_2$] layer and separated by the distance $r$. Because the number of short and long bonds differ, $\chi(r)$ will generally be non-zero even for random distributions of short bonds; consequently we calculate also a statistical correlation function $\chi_{\textrm{rand}}(r)$ which is generated using the same set of $u_i$ values as $\chi(r)$, but with these distributed randomly throughout the configuration. Then the difference function
\begin{equation}
\Delta\chi(r)=\chi(r)-\chi_{\textrm{rand}}(r)
\end{equation}
measures the extent to which short Cu2--O4 bonds are more ($\Delta\chi(r)>0$) or less ($\Delta\chi(r)<0$) likely to cluster at a distance $r$ than in a random distribution.

Calculation of $\Delta\chi(r)$ for any one of our RMC configurations yields a correlation function of the general form shown in Fig.~\ref{fig6}. In all cases we find only one consistently well-defined peak: namely a positive value of $\Delta\chi(r)$ at $r=a_{\textrm{eff}}$ ($=\sqrt{ab}$). This would suggest that, in general, short Cu2--O4 bonds are more likely to cluster into pairs than to be distributed statistically throughout the configuration. If accompanied by charge localisation, then such a result could be considered consistent with a bipolaron model.\cite{Alexandrov_2011}

In conclusion, RMC refinements of neutron PDF data collected for a YBa$_2$Cu$_3$O$_{6.93}$ sample yield atomistic configurations that are simultaneously consistent with all features of the short-range and long-range structural features captured by total scattering experiments and with the results of independent X-ray absorption spectroscopy experiments. We find no evidence to support the use of a split site model, but do find that the apical Cu2--O4 bonds are of two types: a short bond of 2.197(10)\,\AA\ and a long bond of 2.306(3)\,\AA. There appears to be some clustering of the short Cu2--O4 bonds into nearest-neighbour pairs within [CuO$_2$] layers.

\section*{Acknowledgements}

The authors are pleased to acknowledge valuable discussions with R L McGreevy. This research was supported financially by the EPSRC (grant EP/G004528/2) and the ERC (project 279705), and by the STFC in the form of access to the GEM instrument at ISIS.


\begin{references}

\bibitem{Newns_2007}
D. M. Newns and C. C. Tsuei, Nature Phys. {\bf 3}, 184 (2007)

\bibitem{Reznik_2006}
D. Reznik, L. Pintschovius, M. Ito, S. Iikubo, M. Sato, H. Goka, M. Fujita, K. Yamada, G. D. Gu and J. M. Tranquada, Nature {\bf 440}, 1170 (2006)

\bibitem{Orenstein_2000}
J. Orenstein and A. J. Millis, Science {\bf 288}, 468 (2000)

\bibitem{Gadermaier_2010}
C. Gadermaier, A. S. Alexandrov, V. V. Kabanov, P. Kusar, T. Mertelj, X. Yao, C. Manzoni, D. Brida, G. Cerullo and D. Mihailovic, Phys. Rev. Lett. {\bf 105}, 257001 (2010)

\bibitem{Tranquada_1995}
J. M. Tranquada, B. J. Sternlieb, J. D. Axe, Y. Nakamura and S. Uchida, Nature {\bf 375}, 561 (1995)

\bibitem{Alexandrov_2011}
A. S. Alexandrov, Phys. Scr. {\bf 83}, 038301 (2011)

\bibitem{Jin_2007}
C. Q. Jin, Q. Q. Liu, H. Yang, L. X. Yang, R. C. Yu and F. Y. Li, Physica C {\bf 460-462}, 178 (2007)

\bibitem{Conradson_1989}
S. D. Conradson and I. D. Raistrick, Science {\bf 243}, 4896 (1989)

\bibitem{Mustredeleon_1990}
J. Mustre de Leon, S. D. Conradson, I. Bati{\v s}ti{\'c} and A. R. Bishop, Phys. Rev. Lett. {\bf 65}, 1675 (1990).

\bibitem{Mustredeleon_1992}
J. Mustre de Leon, S. D. Conradson, I. Bati{\v s}ti{\'c}, A. R. Bishop, I. D.
  Raistrick, M. C. Aronson and F. H. Garzon, Phys. Rev. B {\bf 45}, 2447 (1992)

\bibitem{Stern_1993}
E. A. Stern, M. Qian, Y. Yacoby, S. M. Heald and H. Maeda, Physica C:
  Superconduct. {\bf 209}, 331 (1993)

\bibitem{Booth_1996}
C. H. Booth, F. Bridges, J. B. Boyce, T. Claeson, B. M. Lairson, R. Liang and
  D. A. Bonn, Phys. Rev. B {\bf 54}, 9542 (1996)
  
\bibitem{Tyson_1997}
T. A. Tyson, J. F. Federici, D. Chew, A. R. Bishop, L. Furenlid, W. Savin and W. Wilber, Physica C {\bf 292}, 163 (1997)

\bibitem{Sullivan_1993}
J. D. Sullivan, P. Bordet, M. Marezio, K. Takenaka and S. Uchida, Phys. Rev. B {\bf 48}, 10638 (1993)

\bibitem{Schweiss_1994}
P. Schweiss, W. Reichardt, M. Braden, G. Collin, G. Heger, H. Claus and A. Erb, Phys. Rev. B {\bf 49}, 1387 (1994)

\bibitem{Francois_1988}
M. Fran{\c c}ois, A. Junod, K. Yvon, A. W. Hewat, J. J. Capponi, P. Strobel, M. Marezio and P. Fischer, Solid State Commun. {\bf 66}, 1117 (1988)

\bibitem{Williams_1988}
A. Williams, G. H. Kwei, R. B. Von Dreele, A. C. Larsen, I. D. Raistrick and D. L. Bish, Phys. Rev. B {\bf 37}, 7960 (1988)

\bibitem{Kwei_1990}
G. H. Kwei, A. C. Larson, W. L. Hults and J. L. Smith, Physica C {\bf 169}, 217 (1990)

\bibitem{Kwei_1991}
G. H. Kwei, A. C. Larson, W. L. Hults and J. L. Smith, Physica C {\bf 175}, 615 (1991)

\bibitem{Louca_1999}
D. Louca, G. H. Kwei, B. Dabrowski and Z. Bukowski, Phys. Rev. B {\bf 60}, 7558 (1999)

\bibitem{Gutmann_2000}
M. Gutmann, S. Billinge, E. Brosha and G. Kwei, Phys. Rev. B {\bf 61}, 11762 (2000)

\bibitem{Liechtenstein_1994}
A. I. Liechtenstein, I. I. Mazin, O. K. Anderson and O. Jepsen, Phil. Mag. B {\bf 70}, 643 (1994)

\bibitem{Arai_1994}
M. Arai, K. Yamada, S. Hosoya, A. C. Hannon, Y. Hidaka, A. D. Taylor and
  Y. Endoh, J. Superconduct. {\bf 7}, 415 (1994)

\bibitem{Blank_1988}
D. H. A. Blank, H. Kruidhof and J. Flokstra, J. Phys. D: Appl. Phys. {\bf 21}, 226 (1988)

\bibitem{Sharma_1991}
R. P. Sharma, F. J. Rotella, J. D. Jorgensen and L. E. Rehn, Physica C:
  Superconduct. {\bf 174}, 409 (1991)
  
\bibitem{Williams_1998}
W. G. Williams, R. M. Ibberson, P. Day and J. E. Enderby, Physica B: Condens. Matt. {\bf 241-243}, 234 (1997)

\bibitem{Day_2004}
P. Day, J. Enderby, W. Williams, L. Chapon, A. Hannon, P. Radaelli and
  A. Soper, Neutron News {\bf 15}, 19 (2004)

\bibitem{Hannon_2005}
A. C. Hannon, Nucl. Instrum. Methods Phys. Res. A {\bf 551}, 88 (2005)

\bibitem{Dove_2002}
M. T. Dove, M. G. Tucker and D. A. Keen, Eur. J. Miner. {\bf 14}, 331 (2002)

\bibitem{Keen_2001}
D. A. Keen, J. Appl. Crystallogr. {\bf 34}, 172 (2001)

\bibitem{GSAS}
A. C. Larson and R. B. Von Dreele, {\it Los Alamos National Laboratory
  Report LAUR 86-748} (1994)
  
\bibitem{Capponi_1987}
J. Capponi, C. Chaillout, A. Hewat, P. Lejay, M. Marezio, N. Nguyen, B. Raveau,
  J. Soubeyroux, J. Tholence and R. Tournier, Europhys. Lett. {\bf 3}, 1301 (1987)

\bibitem{Tucker_2007}
M. G. Tucker, D. A. Keen, M. T. Dove, A. L. Goodwin and Q. Hui, J. Phys.:
  Condens. Matt. {\bf 19}, 335218 (2007)

\bibitem{Goodwin_2005}
A. L. Goodwin, M. G. Tucker, E. R. Cope, M. T. Dove and D. A. Keen, Phys.
  Rev. B {\bf 72}, 214304 (2005)
  
\bibitem{Farrow_2007}
C. L. Farrow, P. Juhas, J. W. Liu, D. Bryndin, E. S. Bozin, J. Bloch, T. Proffen and S. J. L. Billinge, J. Phys.: Condens. Matt. {\bf 19}, 335219 (2007)

\bibitem{Bozin_2007}
E. S. Bozin, M. Schmidt, A. J. DeConninck, G. Paglia, J. F. Mitchell, T. Chatterji, P. G. Radaelli, T. Proffen and S. J. L. Billinge, Phys. Rev. Lett. {\bf 98}, 429 (2007)

\bibitem{uijnote}
We note that our Rietveld refinement gives the root-mean-squared displacement of the Cu2 and O4 atoms in the $z$ direction to be 0.066 and 0.070\,\AA, respectively; consequently displacements of $\frac{1}{2}\Delta d($Cu2--O4$)\simeq0.055$\,\AA\ away from each site would be indistinguishable from thermal motion.

\bibitem{Alexandrov_1998}
See the comment by R. L. McGreevy in A. S. Alexandrov, Phil. Trans. R. Soc. Lond. A {\bf 356}, 197 (1998).

\end{references}
\end{document}